\documentstyle[preprint,aps,epsf]{revtex}
\begin{document}
\draft
\title{ Determination of nucleation rates near the 
 critical point}
\author{Larissa V. Bravina\footnote{Alexander von Humboldt
Foundation Fellow}~ and Eugene E. Zabrodin}
\address{
Department of Physics, University of Bergen, 
All\'egaten 55, N-5007 Bergen, Norway \\
Institute for Theoretical Physics, University of Frankfurt,\\
Robert-Mayer-Strasse 8-10, D-60054 Frankfurt, Germany \\
Institute for Nuclear Physics, Moscow State University,
119899 Moscow, Russia \\
}

\maketitle

\begin{abstract}
The nucleation rates derived for the condensation from a
supersaturated vapor are examined both in the classical theory
and in the modern coarse-grained field theory. By virtue of the 
scaling variable $\lambda_Z$ it is shown that the method of
steepest descent is irrelevant to evaluate the nucleation 
rate in the proximity of the critical point in the capillary
approximation. If the logarithmic corrections to the activation
energy of a droplet are taken into account, then the calculated
nucleation rates provide an adequate description of the liquid-gas 
phase transition both near and out of the critical range.
\end{abstract}

\newpage

The problem of homogeneous nucleation has been intensively 
investigated both theoretically and experimentally (see, e.g.,
\cite{GMS83,Kelt91,Wu96} and references therein). The first
theoretical approach to it, which is often referred to as classical
nucleation theory, was worked out about 50 years ago mainly by 
Becker and D\"{o}ring \cite{BeDo35} and Zeldovich \cite{Zeld43}.
Later on the classical nucleation theory was generalized to a 
system of arbitrarily many degrees of freedom by Landauer and
Swanson \cite{LaSw61} and to the field theories by Cahn and
Hilliard \cite{CaHi59} and Langer \cite{Lang67,Lang69}. Langer
pointed out that the coarse-grained procedure
can be efficiently performed near the critical point where the
radius of critical fluctuations is larger than the characteristic
correlation length.

Both classical and modern coarse-grained field theory assumes that
the decay of an initially homogeneous metastable state should proceed
via the formation of nucleated clusters of a new, stable state.
The rate of relaxation of the metastable state, or nucleation rate,
is given by the formula 
\begin{equation}
\displaystyle
I = I_0\, \exp{\left( - \beta \Delta F_c \right)}\ .
\label{1}
\end{equation}
Here $I_0$ is the preexponential factor, $\beta \equiv (k_B T)^{-1}$,
$k_B$ is Boltzmann's constant, $T$ is the temperature of the system,
and $\Delta F_c$ is the excess free energy of the critically large 
cluster in the system. Even though the prefactors in both theories have 
different expressions, the numerical results for condensation from 
the supersaturated vapor \cite{LaTu73} are very similar. In principle,
it is possible to show \cite{BrZa97} that under the certain 
assumptions the prefactor derived in the classical theory may be 
obtained in a form equivalent to that of the field theory. To test 
the predictions of the nucleation theories a number of experiments on 
the separation of binary fluids near the critical point
\cite{SuOr62,HeCa73} have been performed. It turned out that 
the critical 
systems were more stable than it was expected from the theoretical
calculations. To explain this difference between theory and experiment
Binder and Stauffer \cite{BiSt76,Bind87} argued that since the 
experimentalists measured the completion time of the transition,
one has to consider both droplet formation and droplet evolution,
growth and shrinkage, during the relaxation process. Near the critical 
point droplets grow very slowly and this circumstance decelerates the
completion of the phase transition.
A detailed description of the nucleation kinetics in near-critical 
fluids similar to the theory of coagulation by Lifshitz and Slyozov
\cite{LiSl61,LiPi81} has been developed by Langer and Schwartz
\cite{LaSch80}. The overall reaction rate was found to be much lower 
than the nucleation rate alone. 

In the present paper we would like to show that the nucleation rate 
itself calculated in the capillary approximation near the critical 
point is overestimated. Therefore, the agreement between theory and 
experiment should be even better. 
For the sake of simplicity, we will consider first the application of 
the classical nucleation theory to a first-order phase transition with
only one order parameter, for instance, condensation from the 
supersaturated vapor.

The excess of the Helmholtz free energy due to the formation of a
spherical droplet of radius $R$ is equal in the capillary (or thin
wall) approximation \cite{LiPi81} to the sum of the bulk and the 
surface energies
\begin{equation}
\displaystyle
\Delta F(R) = - \frac{4\pi}{3}R^3\Delta p \, +\, 4\pi R^2 \sigma \  ,
\label{2}
\end{equation}
containing the difference in pressures inside and outside the 
droplet $\Delta p$ and the surface tension $\sigma$.
Minimization of $\Delta F$ with respect to the radius yields the 
energy of the critically large droplet
$\displaystyle \Delta F_c = \frac{4}{3} \pi R_c^2 \sigma$, where
the critical radius is given by Laplace's formula 
$R_c = 2 \sigma / \Delta p$. Using the reduced variables
\cite{BrZa95} $\displaystyle \lambda_Z = R_c \, \sqrt{ 4 \pi \sigma 
\beta } $ and $r = R/\!R_c$, we have
\begin{equation}
\displaystyle
\beta \Delta F = - \frac{2}{3} \lambda_Z^2 r^3 + \lambda_Z^2 r^2
\label{3}
\end{equation}
and
\begin{equation}
\displaystyle
\beta \Delta F_c = \frac{1}{3} \lambda_Z^2 \ .
\label{4}
\end{equation}

As first shown by Zeldovich \cite{Zeld43}, the size distribution
function of droplets $f(R)$ obeys the Fokker-Planck kinetic equation 
\begin{equation}
\displaystyle
\frac{\partial f}{\partial t} = - \frac{\partial J}{\partial R}\ ,
\label{5}
\end{equation}
where
\begin{equation}
\displaystyle
J = -B \frac{\partial f}{\partial R} + A f
\label{6}
\end{equation}
is a so-called current in the size space, containing the diffusion
and the drift coefficients $B$ and $A$. The solution of Eq.~(\ref{5}) 
corresponding to $J = J_0 = 0$ is a well-known canonical distribution 
$f_0(R) \propto \exp{(-\beta \Delta F)}$. In terms of the ratio 
$f(R)/f_0(R)$, Eq.~(\ref{6}) may be rewritten as
\begin{equation}
\displaystyle
J = B f_0(R) \frac{\partial}{\partial R} \left( \frac{f(R)}{f_0(R)}
\right)\ .
\label{7}
\end{equation}

The theory of first-order phase transitions is based on the 
requirement of the steady-state flux $J = J_{ss} = {\rm const}$. 
In the region of small $r$ the distribution function $f(R)$ is very 
close to the equilibrium distribution function $f_0(R)$, whereas it 
diminishes quickly for $r \geq 1$. These boundary conditions represent 
the constant replenishment of the stock of small droplets because of 
the thermodynamic fluctuations and the fact that droplets of 
supercritical sizes are removed from the system and merge into the new 
phase. The steady-state solution of Eq.~(\ref{7}) that satisfies both 
boundary conditions reads
\begin{equation}
\displaystyle
J_{ss}^{-1} = \int_0^\infty \frac{d R}{B f_0(R)}\ .
\label{8}
\end{equation}
In order to evaluate the integral analytically it is usually mentioned
that the integrand has a very sharp maximum at $R = R_c$ due to the
maximum of the activation droplet energy at this point. Therefore, one
may replace the activation energy near the critical radius by its
harmonic approximation
\begin{eqnarray}
\displaystyle
\beta \Delta F &=& \beta (\Delta F )_{R=R_c} +
\frac{1}{2} \beta \left( \frac{\partial^2 \Delta F}{\partial R^2}
\right)_{R=R_c} (R - R_c)^2 \\ 
\nonumber
 &=& \beta \Delta F_c - \lambda_Z^2 (r - 1)^2
\label{9}
\end{eqnarray}
and then apply the method of steepest descent.
We have
\begin{equation}
\displaystyle
J_{ss} = \frac{B(r_c)\, f_0(r_c)}{R_c}\, \sqrt{ -
\frac{\beta}{2 \pi} \left(\frac{\partial^2 \Delta F}{\partial r^2} 
\right)_{r = r_c} }  
= \frac{B(1) f_0(1)}{R_c}\, \frac{\lambda_Z}{ \sqrt{\pi} }\ .
\label{10}
\end{equation}
Mathematically, the assumption that $\Delta F(r)$ has a sharp maximum
at $r_c = 1$ means that
\begin{equation}
\displaystyle
\left[ \sqrt{\beta |\Delta F^{\prime \prime} (1) | } \right]^{-1}
\ll 1 \ ,
\label{11}
\end{equation}
where $\Delta F^{\prime \prime}$ denotes the second derivative of
$\Delta F$ with respect to the reduced radius $r$.

It is easy to see that in the capillary approximation
\begin{equation}
\beta |\Delta F^{\prime \prime}(1)| = 2 \lambda_Z^2\ .
\label{12}
\end{equation}
Our next step is to determine the dependence of the similarity number
$\lambda_Z$ on the critical exponents provided the temperature of 
the system approaches the critical one.

As functions of $\theta = 1 - T/T_c$, the thermodynamic quantities 
needed for our analysis have the following power-law approximations
in the vicinity of the critical point:
\begin{equation}
\sigma \sim \theta^{\mu^\prime}\ \ \ ,\ \ \ \Delta p \sim \theta^{
\beta^\prime \delta^\prime}\ .
\label{13}
\end{equation}
In classical theory the critical exponents $\beta^\prime$ and 
$\delta^\prime$ are equal to 1/2 and 3, respectively. The measured
value of the exponent $\mu^\prime$ lies in the range 1.22$-$1.29
\cite{surft} and we will use $\mu^\prime = 1.25$ in our further
calculations. Thus the critical radius tends to infinity at a 
critical temperature as 
$\theta^{(\mu^\prime - \beta^\prime \delta^\prime)} = \theta^{-1/4}$. 
The similarity number near the critical temperature obeys the power law 
\begin{equation}
\displaystyle
\lambda_Z = R_c \sqrt{4 \pi \sigma \beta} \sim 
\theta^{(\frac{3\mu^\prime}{2} - \beta^\prime \delta^\prime)} =
\theta^{3/8}
\label{14}
\end{equation}
It is clear that the parameter $\lambda_Z $ tends to zero at 
$T \rightarrow T_c$ and that the criterion (\ref{11}) is not fulfilled.
Therefore, in the thin wall approximation the method of steepest 
descent is not applicable to the evaluation of the nucleation rate 
near the critical point.

In the semiphenomenological droplet model worked out by Fisher 
\cite{Fish67} the activation free energy includes also the curvature 
term related to the small fluctuations in the shape of the droplet 
which do not change both the volume and the surface area of the 
droplet
\begin{equation}
\displaystyle
\Delta F^F(R) = -\frac{4}{3} \pi R^3 \Delta p + 4 \pi \sigma R^2
 + 3 \tau \beta^{-1} \ln{\frac{R}{r_0}} \ .
\label{15}
\end{equation}
Here $\tau$ is the Fisher critical exponent, which is about 2.2, and
$r_0$ is the radius of the smallest droplet in the system. 
It is worth noting that the spherical harmonic excitations of a
droplet (Goldstone modes) have been calculated \cite{Lang67,GNW80} 
also in the field theory. These calculations lead to the appearance 
of the logarithmic term in the expansion of the free-energy density,
similar to the curvature term in Eq.~(\ref{15}), with the critical
exponent $\tau = 7/3$. 
Then in the harmonic approximation
\begin{equation}
\displaystyle
\beta \Delta F^F = \beta \Delta F^F_c - \frac{9 \tau + 2 \lambda_Z^2}
{2}\, (r - 1)^2 \ ,
\label{16}
\end{equation}
\begin{equation}
\displaystyle
\beta \Delta F^F_c = -\tau + \frac{\lambda_Z^2}{3} + 3 \tau 
\ln{\frac{R_c^F}{r_0} } \ .
\label{17}
\end{equation}
Note that the critical radius $R_c^F$ appearing in Eq.~(\ref{17}) 
is not the same as that given by Laplace's formula, but should be 
determined by the solution of cubic (with respect to $R$) equation. 
Now 
\begin{equation}
\displaystyle
\beta |[\Delta F^F(1)]^{\prime \prime}| = 9 \tau + 2 \lambda_Z^2
\label{18}
\end{equation}
and the method of steepest descent is relevant since the criterion
(\ref{11}) is fulfilled:
\begin{equation}
\displaystyle
\left( 9 \tau + 2 \lambda_Z^2 \right)_{\lambda_Z \rightarrow 0}^{-1} 
\longrightarrow (9 \tau)^{-1} \ll 1 \ .
\label{19}
\end{equation}
Performing the saddle-point integration in Eq.~(\ref{8}), one can 
find the nucleation rate in the droplet model approach
\begin{equation}
\displaystyle
J_{ss}^F = \frac{B^F(1) f_0^F(1)}{R_c^F}\, \sqrt{ \frac{ 9 \tau + 2 
\lambda_Z^2}{ 2 \pi} }\ .
\label{20}
\end{equation}

\begin{figure}[htp]
\centerline{\epsfysize=15cm \epsfbox{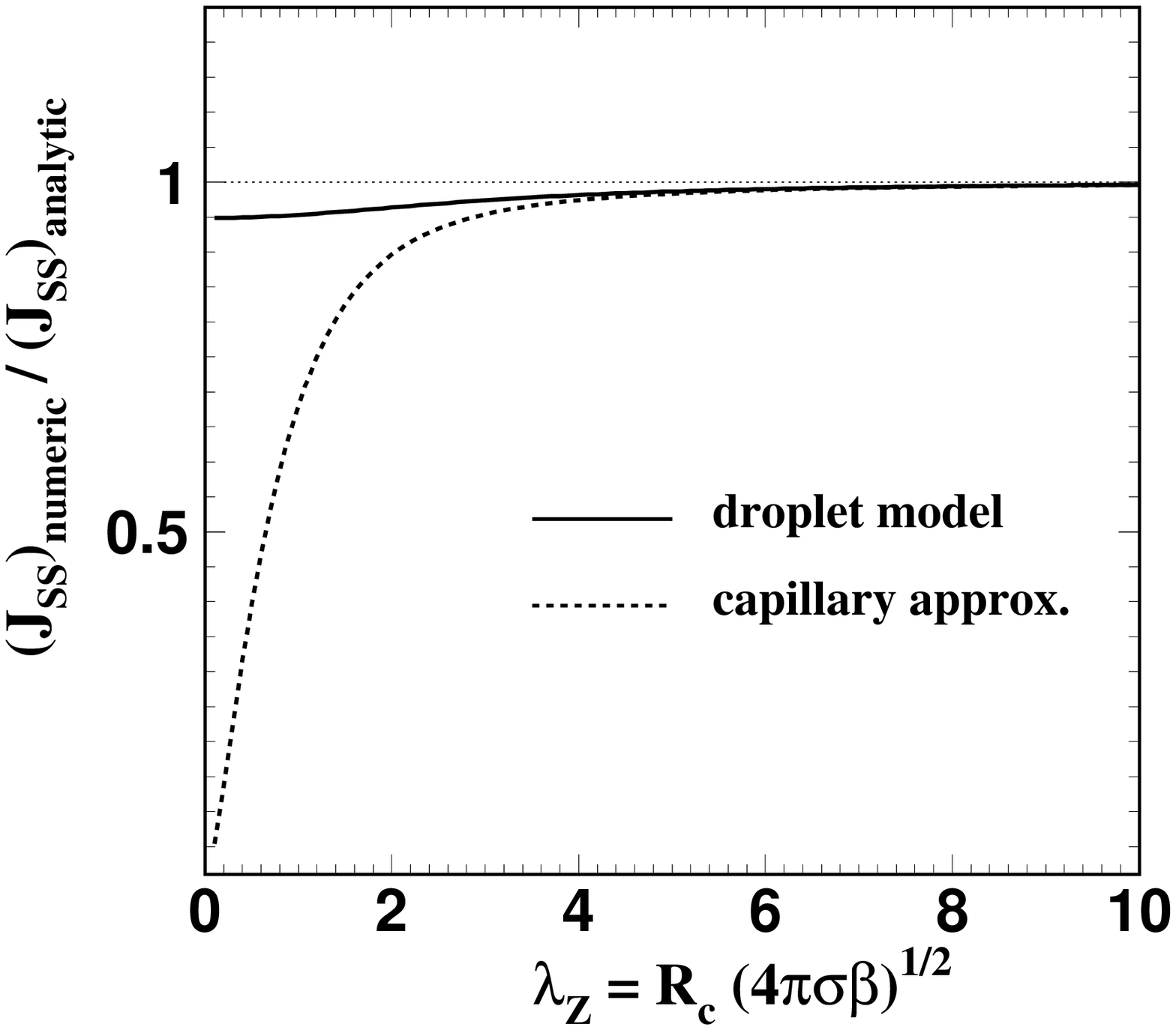}}
\caption{ 
Ratio $\displaystyle \left( J_\protect{ ss \protect} \right)_
\protect{ \rm numeric \protect}/ \left( J_\protect{ ss \protect} 
\right)_\protect{ \rm analytic \protect} $ versus the parameter 
$\displaystyle \lambda_Z = R_c \left( 4 \pi \sigma \beta 
\right)^{1/2}$
corresponding to the Fisher droplet model (solid line)
and to the capillary approximation (dashed line).}

\label{f1}
\end{figure}

To compare the analytical expressions given by Eqs.~(\ref{10}) and
(\ref{20}) with the numerical solutions of Eq.~(\ref{8}), 
we plot in Fig. \ref{f1} the ratio $\displaystyle \left( J_{ss} 
\right)_{\rm numeric}/\left( J_{ss} \right)_{\rm analytic}$ versus 
$\displaystyle \lambda_Z = R_c\, \sqrt{4 \pi \sigma \beta}$
calculated both in the capillary approximation and in the Fisher 
droplet model. We see that formula (\ref{20}) gives us values of 
the nucleation rate $J_{ss}$ that agree with the results of 
numerical calculations by Eq.~(\ref{8}) within the $5\%$ accuracy 
limit even for very small values of $\lambda_Z$. In contrast, in 
the capillary approximation presented by Eq.~(\ref{10}), significant
deviations from the numerical results start already at $\lambda_Z=2$,
which is assigned to a system rather far from the critical point.
Thus the value $\lambda_Z=2$ may be considered as a limit of 
applicability of the classical expression (\ref{10}) for the 
nucleation rate. At larger values of the similarity number both
analytical expressions (\ref{10}) and (\ref{20}) fit well to the
results of numerical calculations. 

The errors introduced by the method of steepest descent have been
calculated numerically \cite{Cohe70} for the condensation of a gas.
It was found that the errors are negligibly small except for the
smallest critical droplets and for the smallest values of the 
activation energy of the critical droplet. We show that this error
can be parametrized by the single scaling variable $\lambda_Z$.
Our analysis is valid for the Langer theory \cite{Lang69} also, in 
which the prefactor $I_0$ is shown to be a product of the dynamical 
and statistical prefactors $\kappa$ and $\Omega_0$, respectively,
\begin{equation}
\displaystyle
I_0 = \frac{|\kappa|}{2 \pi}\, \Omega_0\ .
\label{21}
\end{equation}
The dynamical prefactor describes the exponential growth rate of the
unstable mode at the saddle point. It is related to the single 
negative eigenvalue $\lambda_1$ of the generalized mobility matrix
$M_{i j} = \partial^2 \Delta F/\partial \xi_i \partial \xi_j $, 
where $\{\xi_i\}$ is a set of macroscopic variables
describing the system. The integration over a plane containing the
saddle point is performed by the method of steepest descent and the 
statistical prefactor becomes
\begin{equation}
\displaystyle
\Omega_0 = {\cal V}\  \left( \frac{2 \pi}{\beta |\lambda_1|} \right)
^{1/2}\ \left[ \frac{{\rm det}(\beta M_0 / 2 \pi)}{{\rm det}
(\beta M^\prime /2 \pi)} \right]^{1/2}\ ,
\label{22}
\end{equation}
where ${\cal V}$ is the available phase-space volume of the saddle
point, the index 0 denotes the metastable state, and a prime 
indicates that the negative eigenvalue $\lambda_1$
as well as the zero eigenvalues of the matrix $M_{ij}$ is omitted.
For the process of vapor condensation the simplified model contains
only the order parameter. From the definition of the mobility matrix
it follows that $ - \beta \lambda_1 = 2 \lambda_Z^2$ 
[cf. Eq.~(\ref{12})]. Therefore, Eq.~(\ref{22}) cannot be applied in 
the capillary approximation for the liquid-gas system near the 
critical point because the criterion (\ref{11}) is violated. 
To calculate the nucleation rate in the critical region one has 
either to evaluate the integral numerically or to insert the
logarithmic corrections \cite{Lang67,GNW80} mentioned above in the
activation energy of a droplet {\it before\/} the steepest-descent
evaluation of the integral. In the latter case this curvature term
will play a crucial role in the determination of the saddle point
for the free-energy functional.
This is the last important point in our discussion. 
Equation (\ref{1}) can be rewritten also \cite{Lang67} via the 
imaginary part Im$\,{\cal F}$ of the analytic continuation of the
free-energy density to the metastable state 
\begin{equation}
\displaystyle
I = \frac{\beta |\kappa|}{\pi}\, {\rm Im}\,{\cal F}\ ,
\label{23}
\end{equation}
and the evaluation of Im$\,{\cal F}$ has been widely discussed in the
literature. Sometimes Im$\,{\cal F}$ has a form \cite{LaSch80} like
$x^\tau \exp{(-x^2)}$, where the dimensionless variable $x^2$ 
corresponds to the activation energy of a critical droplet times 
$\beta$, and the saddle point is calculated {\it in the capillary
approximation\/}. In the context of a steepest descent this implies 
that the term with the logarithmic
corrections $x^\tau$ is considered as a slowly varying part of the 
integrand in the vicinity of saddle point. Therefore, these 
corrections are added to the activation energy {\it after\/} the 
saddle-point evaluation of the integral and the nucleation rate is 
overestimated again near the critical point.

We would like to thank the Department of Physics, University of 
Bergen and the Institute for Theoretical Physics, University of
Frankfurt for the warm and kind hospitality. L.B. acknowledges 
support of the Alexander-von-Humboldt Foundation.

\end{document}